\documentclass[12pt]{article}

\usepackage{amsfonts}
\usepackage{amssymb}
\usepackage{amsmath}

\usepackage{graphicx}
\usepackage{latexsym}
\usepackage{color}
\usepackage[colorlinks=true]{hyperref}
\usepackage{authblk}

\usepackage{graphicx,tikz}
\usepackage{amsmath,amssymb,mathrsfs}

\usepackage{amsthm}

\newcommand{\bm}[1]{\mbox{\boldmath $#1$}}

\def\be{\begin{equation}}
\def\ee{\end{equation}}
\def\bea{\begin{eqnarray}}
\def\eea{\end{eqnarray}}
\def\bean{\begin{eqnarray*}}
\def\eean{\end{eqnarray*}}

{\theoremstyle{definition}
}

\def\scri{\mathscr{J}}

\newlength{\cellwidth}
\begin{document}
\title{Ultra-massive spacetimes in 2+1 dimensions with positive $\Lambda$}

\author{Ingemar Bengtsson$^1$ and Jos\'e M. M. Senovilla$^{2,3}$}
\affil{$^1$ Stockholms Universitet, AlbaNova, SE-106 91 Stockholm, Sweden\\
$^2$ Departamento de F\'isica, Universidad del Pa\'is Vasco UPV/EHU, Apartado 644, 48080 Bilbao, Spain\\
$^3$ EHU Quantum Center, Universidad del Pa\'{\i}s Vasco UPV/EHU.\\}

\


{\let\newpage\relax\maketitle}

\vspace{-0.2em}

\begin{abstract}
We show that ultra-massive spacetimes exist also in 2 + 1 dimensions with a positive cosmological constant $\Lambda >0$.  They can be created through the collapse of a spherical null dust shell. The exterior of the shell is then a Mess spacetime, that is to say a locally de Sitter spacetime that cannot be obtained as a quotient of de Sitter space. 
\end{abstract}

\section{Introduction}
In standard 4-dimensional General Relativity (GR) with a positive cosmological constant $\Lambda >0$ there is a bound for the area of spatially stable marginally trapped surfaces given by $4\pi/\Lambda$ \cite{HSN,Simon,W}. As largely discussed in \cite{Snew,Snew2} this does not lead to absence of ultra-strong gravity collapsed objects, but rather to the existence of a new kind of extremely powerful collapsed objects---called `ultra-massive'---that show gravitational properties even more powerful than those of black holes. In particular, future null infinity does not exist being replaced by a future singularity, hence there is no concept of event horizon but there is a marginally trapped tube (MTT) (see \cite{AK,BeS,Booth,S} for definitions) with very peculiar properties \cite{Snew,Snew2}.

In this paper we analyze the question of whether similar objects may exist in 2+1 GR with $\Lambda >0$. As is well known, there are no black-hole solutions in this case. By contrast, when $\Lambda$ is negative there are black hole solutions \cite{BTZ}, as well as remarkable analogies between 
these BTZ black holes and 3+1 dimensional black hole physics \cite{Carlip}. For positive $\Lambda$ we will use simple models with a collapsing finite shell of spherical null dust to prove that  `ultra-massive' spacetimes do arise in 2+1 dimensions with $\Lambda >0$---along with other possibilities leading to naked singularities. The ultra-massive models possess a future singularity of Misner type \cite{M,S1} as well as a curvature singularity inside the matter. There is also a very special critical case where a remnant of future infinity survives for timelike observers, but otherwise having a null singularity too.

The regions outside matter in the ultra-massive cases are locally de Sitter (dS) but they are {\em not} portions of a 3-dimensional de Sitter spacetime, nor 
are they quotients thereof. That such strange locally dS spacetimes exist was first noticed by Mess \cite{Mess,A}. A simple explanation of how 
they arise has been given \cite{BH}, but we believe that this is the first time they arise in a `natural' problem. It is interesting to compare to the 
case when a spherical null dust shell collapses in 2+1 dimensional anti-de Sitter space. Then, provided the `mass' is large enough, the exterior 
of the shell is given by a portion of a BTZ black hole spacetime \cite{Husain1, Husain2}, and the BTZ black holes are quotients of anti-de Sitter space. In 
a sense then the Mess spacetimes are needed to complete the corresponding picture in de Sitter case. 

For orientation we devote Section 2 to some facts about 2+1 dimensional locally de Sitter spacetimes. Section 3 contains our 
main results, and Section 4 provides some discussion.  

\section{Some expressions for 3-dimensional (locally) de Sitter spacetime}\label{sec:dS}
\begin{figure}[h!]
\includegraphics[width=12cm]{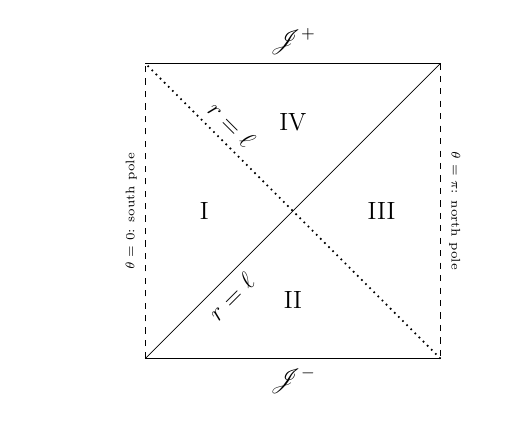}
\caption{{\small A Penrose diagram of the de Sitter 3-dimensional hyperboloid. Here and in the rest of the diagrams null radial geodesics are at 45$^o$ as usual. In the coordinates of \eqref{dSstandard} each point represents a $u=$constant and $\theta=$constant circle coordinatized by $\varphi$ ---or equivalently, by $X$ and $Y$ with $X^2 +Y^2$=constant--- except for the vertical edges where $\cos\theta =\pm 1$ (that is, $X=Y=0$ and $Z^2-U^2 =\ell^2$) which represent the worldlines of the north and south poles in the spheres with constant $u$. $\scri^\mp$ are past and future null infinity, corresponding to $u\rightarrow \mp \infty$.  The lines marked $r = \ell$ are two lightcones with vertices on $\scri^\mp$, given by $Z = \pm U$, and they divide the spacetime into four regions.}\label{fig:deSitter}}
\end{figure}

De Sitter spacetime (dS for short) is usually defined as the hyperboloid
\be
X^2 + Y^2 + Z^2 - U^2 = \ell^2 = \frac{1}{\Lambda} 
\ee
embedded in a one-higher-dimensional Minkowski spacetime with $\{U,X,Y,Z\}$ standard Minkowskian coordinates. Here $\ell$ is a length scale set by the cosmological constant $\Lambda$.\footnote{In what follows, we will use $\Lambda >0$ or $\ell >0$ indistinctly, hoping this will not lead to confusion. Any of the two can be replaced by the other by means of $\Lambda =1/\ell^2$ everywhere.} Using embedding coordinates adapted to the spheres at constant $U$ 
on the hyperboloid, namely,
\be
\begin{array}{l}X = \ell \cosh(u/\ell) \sin\theta\cos{\varphi} \\ Y =\ell \cosh(u/\ell) \sin\theta \sin{\varphi} \\ Z = \ell\cosh(u/\ell) \cos\theta \\ U= \ell\sinh (u/\ell)
\end{array} \hspace{8mm} 
\begin{array}{c} - \infty < u < \infty \\ 0 < \theta < \pi \\ 0 \leq \varphi < 2\pi \ . \end{array}
\ee
one obtains the familiar form of the dS line element
\be\label{dSstandard}
ds^2 = -du^2 + \ell^2 \cosh^2(u/\ell) \left(d\theta^2 +\sin^2\theta d\varphi^2 \right) .
\ee
To orient ourselves we give a Penrose diagram for de Sitter spacetime, divided into four regions depending on the values taken by 
the embedding coordinates, see figure \ref{fig:deSitter}. For purposes of visualisation a different global coordinate system, making 
use of a conformal embedding in the static Einstein universe, is often preferable \cite{BH}. 

The coordinates that we will use in section \ref{sec:models} cover `one half' of de Sitter spacetime ---namely regions I and II in the Penrose diagram of figure \ref{fig:deSitter}---and are given by 
\be
Z > U:  \hspace{8mm} \begin{array}{l}X = r\cos{\varphi} \\ Y = r\sin{\varphi} \\ Z+U = e^{v/\ell}(\ell - r) \\ Z-U = e^{-v/\ell}(\ell + r) 
\end{array} \hspace{8mm} 
\begin{array}{c} - \infty < v < \infty \\ r > 0 \\ 0 \leq \varphi < 2\pi \ . \end{array}
\ee
We find 
\be
ds^2 = dX^2+dY^2+dZ^2-dU^2 = - (1-\Lambda r^2)dv^2 + 2dr dv + r^2d\varphi^2 \  . 
\ee
This is on the (advanced) Eddington-Finkelstein form. 

Another coordinate system that can be made to cover the same region is that of the `steady-state' coordinates, 
\be
Z > U:  \hspace{8mm} \begin{array}{l} X = \displaystyle{\frac{\ell x}{\tilde t}} \\ \displaystyle{Y = \frac{\ell y}{\tilde t}} \\ \displaystyle{Z+U = \tilde t - \frac{x^2+y^2}{\tilde t}}  \\ \displaystyle{Z-U = \frac{\ell^2}{\tilde t} }
\end{array} \hspace{8mm} 
\begin{array}{c} \tilde t > 0 \\ - \infty <  x, y < \infty  \end{array}
\ee
leading to
\be
ds^2 =\frac{\ell^2}{\tilde t^2}(-d\tilde{t}^2 + dx^2 + dy^2) \  . 
\ee
A reparametrisation of the time coordinate $\tilde t = \ell e^{\tau/\ell}$ takes us to the metric (\ref{metric2}), which will play a critical role for us. These coordinates are 
adapted to a pair of `null screws' in the embedding space. It pays to express the manifest Killing vectors in embedding coordinates,  
\be
\ell \partial_y = (Z-U)\partial_Y - Y(\partial_Z + \partial_U) \ , \hspace{6mm} || \partial_y||^2 = (Z-U)^2/\ell^2 \  , 
\ee
and similarly for $\partial_x$. Surfaces at constant $\tilde t$ can be turned into cylinders by making the coordinate $y$ periodic: $y\rightarrow \ell \varphi$. As a result there 
will be closed null curves when $Z = U$ (where the coordinate system ends), and a null singularity of Misner type where $Y = 0$ and the Killing vector 
vanishes. The null singularity consists of two portions, for which $X = \pm \ell$. In covering space they are two generators of the null cone $Z = U$ meeting at a single point on $\scri^+$, the vertex of the null cone. Further details will be given in Section \ref{dScases}. 

To see a Mess spacetime arising we focus on region II in the Penrose diagram, in popular parlance on a `contracting region' of the hyperboloid, and 
introduce the coordinates 
\be
 U < - |Z|:  \hspace{8mm} \begin{array}{l}X = \sqrt{\ell^2+\hat t^2}\cos{z} \\ Y = \sqrt{\ell^2 + \hat t^2}\sin{z} \\ Z = -\hat t\sinh{\varphi} \\ 
 U = -\hat t\cosh{\varphi} \end{array} \hspace{8mm} 
\begin{array}{c} - \infty < \hat t < \infty \\ -\infty < \varphi < \infty \\ 0 \leq z < 2\pi \ . \end{array}
\ee
Note that the coordinate $z$ is periodic, while $\varphi$ is not. We receive
\be
ds^2 = - \frac{d\hat{t}^2}{1 + \Lambda \hat{t}^2}  + \hat{t} ^2d\varphi^2  + (\ell^2 +\hat t^2)dz^2\  . 
\ee
We can introduce a new time coordinate $T$, so that we obtain the local metric in the form \eqref{metric1} we use in Section 
\ref{sec:models}: 
\be
\hat t = \ell \sinh{(T /\ell)} \hspace{5mm} \Rightarrow \hspace{5mm} ds^2 = - dT^2 + \ell^2 \cosh^2{(T/\ell )}dz^2 + \ell^2\sinh^2{( T/\ell )}d\varphi^2 \ . 
\ee
What we have here is a region of de Sitter spacetime foliated by infinitely long cylinders that touch past infinity at two antipodal points, where $\varphi \rightarrow \pm \infty$. 
But we can unroll these cylinders by going to a covering space, so that $- \infty < z < \infty$. The result is a locally de Sitter 
spacetime that cannot be fitted into de Sitter spacetime in any way. It is a perfectly sensible spacetime in itself, in particular the Cauchy development of a surface at constant $T$ is complete towards the past. 
For an illustration, see Figure 7 in ref. \cite{BH}.

There are two manifest Killing vectors. In embedding coordinates they are 
\be 
\partial_z = X\partial_Y - Y\partial_X \ , \hspace{8mm} \partial_\varphi = U\partial_Z +Z\partial_U \ . 
\ee 
It is seen that the Killing vector field $\partial_\varphi$ has fixed points on $U = Z = 0$, which is the equator of a round sphere to the future 
of the coordinate region. In Section 
\ref{sec:models} we will turn $\varphi$ into a periodic coordinate, and this will give rise to a Misner singularity at those fixed points. 

As a final remark we note that the Mess spacetime is not a $2+1$ dimensional pathology. Similar examples can be constructed in 
$3+1$ dimensions \cite{BH}, but they are not of any concern here. 

\section{The models}\label{sec:models}
In local advanced Eddington-Finkelstein-like coordinates $\{v,r,\varphi\}$ the circularly symmetric line-element we are going to study reads
\be\label{metric}
ds^2 = -(F(v) -\Lambda r^2) dv^2 +2 dv dr +r^2 d\varphi^2
\ee
where $\varphi$ is an angular variable (usually $\varphi\in (0,2\pi)$) describing the preferred circles of symmetry, $r>0$ is a length coordinate such that full round circles (defined by constant values of $v$ and $r$) have length $2\pi r$, and $v$ is a null advanced coordinate with values in the real line. As in the previous section, $\Lambda =\ell^2 >0$ represents a positive cosmological constant and $F(v)$ is a function of $v$. When $F(v) = 1$ we recover de Sitter space. The null vector field $\vec k:= -\partial_r$ is assumed to be future-pointing. 

In analogy with the 4-dimensional Vaidya radiating metric, sometimes the function $F(v)$ is rewritten as 
$$
F(v) = 1-m(v)
$$
and then $m(v)$ is called the mass function. It equals zero in de Sitter space, and will be non-negative in all the solutions we consider. 
The metric \eqref{metric} is a solution of the Einstein field equations 
$$
G_{ij} +\Lambda g_{ij} = T_{ij}
$$
for an energy-momentum tensor of null dust type:
\be\label{emt}
T_{ij} = -\frac{\dot F}{2r} k_i k_j , \hspace{1cm} \bm{k} = -dv
\ee
so that at any region with non-constant $F(v)$ there is a curvature singularity at $r\rightarrow 0$. Furthermore, the dominant energy condition requires $\dot F \leq 0$. 

We recall that a submanifold is called (future) trapped, respectively (future) marginally trapped,  if its mean curvature vector $\vec H$ is (future) timelike, resp.\ (future) null, see e.g. \cite{BeS,GS,MS1,S} and references therein. In our case, the preferred round circles defined by constant values of $v$ and $r$ have the mean curvature vector
$$
\vec H =\partial_v + (F(v) -\Lambda r^2)\partial_r .
$$
Notice that as a one-form this is just $dr$. 
Thus, computing the norm of $\vec H$ we deduce that the round circles 
are future-trapped if $r^2 < F(v)/\Lambda$, and marginally future-trapped if $r^2 =F(v)/\Lambda$. Hence, they can only exist in regions with $F(v)> 0$. In these regions, there is a surface foliated by these marginally trapped circles, that is to say, a circularly symmetric MTT, defined by
\be\label{MTT}
r=\sqrt{\frac{F(v)}{\Lambda}} .
\ee
This MTT is spacelike (a dynamical horizon \cite{AK,S}) whenever $\dot F := dF/dv >0$, timelike  (a timelike membrane \cite{Booth,S}) if  $\dot F < 0$, and it becomes null wherever  $\dot F =0$. In particular the MTT is null in any region of the spacetime which is locally dS ($F$ is constant then). The dominant energy condition then requires that the MTT is non-spacelike everywhere: a timelike membrane with possible null portions.

The surfaces $r=$ const.\ are spacelike if $F(v) <\Lambda r^2$, timelike if $F(v) >\Lambda r^2$ and null at the MTT $F(v) =\Lambda r^2$ if this exists. 

\subsection{Locally dS cases: $F(v)=$ constant}\label{dScases}
The case with $m(v)=0 \Longleftrightarrow F(v)=1$ describes a locally de Sitter spacetime. This is just half of the full dS spacetime, as explained in section \ref{sec:dS}. Actually, every region with $F(v)=$constant is locally isometric to dS, however the regularity of the axis at $r=0$ needs that $F$ be 1. To prove this, note that the axial Killing vector $\partial_\varphi$ has norm $r^2$ and then the regularity condition for the axis \cite{MS,Exact} requires
$$
\lim_{r\rightarrow 0} \, (F(v) -\Lambda r^2) = F(v) =1 .
$$
Therefore, when $F(v)\neq 1$, and in particular if $F(v)$ is constant on a region but different from 1, then there is either a conical singularity at $r=0$ (if $0 < F < 1$) or something stranger if $F\leq 0$, as discussed next. 

Consider first the case with $F=k^2 <1$. Then a trivial re-parameterization of the coordinates
$$
\tilde v = k v, \hspace{5mm} \tilde r = r/k, \hspace{5mm} \tilde\varphi =k\varphi
$$
allows us to rewrite the metric in the original form \eqref{metric} in the new coordinates, but notice that now $\tilde\varphi \in (0,2 k \pi)$. Thus, there is a deficit angle, ergo a conical singularity, quantified by $k$.

Let now $F=-a^2 <0$. In this case the MTT is not present, the surfaces $r=$ constant are all spacelike , so that $r$ is a time coordinate, and all round circles are future-trapped. As shown in section \ref{sec:dS} in this situation the following change of coordinates
$$
z=v+\frac{\ell}{a} \arctan \left(\frac{r}{a\ell} \right), \hspace{1cm} r= -a\ell \sinh\left(\frac{T}{\ell}\right) 
$$
 rewrites the metric in the form
\be\label{metric1}
ds^2 = -dT^2 +a^2\cosh^2\left( \frac{T}{\ell}\right) dz^2 +\ell^2 a^2 \sinh^2\left( \frac{T}{\ell}\right) d\varphi^2
\ee
where $-\infty < T <0$ corresponds to $r$ going from $\infty$ to $0$. There arises a Misner-like singularity \cite{M,S1} at $T\rightarrow 0$, where null geodesics with say $z=$constant are incomplete, as they spiral an infinite number of times without ever reaching $T=0$, and they do that in finite affine parameter.

This type of metric \eqref{metric1} is locally dS but it is {\em not} a portion of full dS. See section \ref{sec:dS} for details of how such `Mess spacetimes' arise.

There is a critical case when $F=0$. Now the following change
$$
x=v+\frac{\ell^2}{r}, \hspace{1cm} r=\ell e^{-\tau/\ell} 
$$
brings the metric to a `steady state' form of dS
\be\label{metric2}
ds^2 =- d\tau^2 + e^{-2 \tau/\ell } \left(dx^2 +\ell^2 d\varphi^2 \right)  
\ee
with $-\infty < \tau <\infty$ as $r$ runs from $\infty$ to $0$. Observe, though, that the surfaces of symmetry $\tau =$ constant are not planes, but cylinders. As $\tau \rightarrow \infty$ there arises a {\em null} singularity, but it has several distinctive features that we will need to know because this type of spacetimes appears in the critical case studied in subsection \ref{subsec:critical}. To decipher the properties of this future singularity we consider the geodesic equations, that can be easily derived leading to
$$
\varphi' =\Lambda L e^{2\sqrt{\Lambda}\tau}, \hspace{1cm} x'= C e^{2\sqrt{\Lambda}\tau}, \hspace{1cm} -\tau'^2 +e^{2\sqrt{\Lambda}\tau}C^2 +\Lambda e^{2\sqrt{\Lambda} \tau} L^2= \epsilon
$$
where prime denotes derivative with respect to the affine parameter $\lambda$, $C$ and $L$ are constants of integration and $\epsilon =-1$ for timelike geodesics while $\epsilon =0$ for null ones.

Consider first the null geodesics with $C=0$. A direct integration provides then 
$$
x= x_0, \hspace{1cm} e^{-\sqrt{\Lambda} \tau} =\pm \Lambda L (\lambda-\lambda_0), \hspace{1cm} \varphi -\varphi_0=\mp \frac{1}{\Lambda L (\lambda-\lambda_0)}
$$
showing again a Misner-like behaviour for they reach $\tau\rightarrow \infty$ in finite affine parameter but spiraling indefinitely. The null geodesics without angular momentum $L=0$, on the other hand, are given by
$$
\varphi =\varphi_0, \hspace{1cm} e^{-\sqrt{\Lambda} \tau} =\pm \sqrt{\Lambda} C (\lambda-\lambda_0), \hspace{1cm} x-x_0 = \pm \frac{1}{C \Lambda (\lambda -\lambda_0)}.
$$
They reach $\tau\rightarrow \infty$ in finite affine parameter but now half of them reach there with $x\rightarrow \infty$ and the other half reach there with $x \rightarrow -\infty$. Indeed, as discussed in Section \ref{sec:dS}, the singularity consists of two disconnected parts, meeting at $\scri$. 
One of them shows the endpoints of the `outgoing' geodesics, the other the endpoints of `ingoing' geodesics. 

This can be further supported by considering the timelike geodesics with $L=C=0$. They are described by
$$
x=x_0, \hspace{1cm} \varphi =\varphi_0, \hspace{1cm} \tau -\tau_0 =\lambda -\lambda_0 .
$$
As we see, these timelike geodesics are {\em complete} and they only arrive at $\tau \rightarrow \infty$ after an infinite proper time has elapsed. Therefore, there is future infinity in this spacetime, even though no null geodesic can reach there. This future timelike infinity separates the two previous non-curvature singularities where the null geodesics arrive in finite affine parameter. Therefore, this structure will arise in the critical case of subsection \ref{subsec:critical}.

\subsection{The collapsing finite shell} 
We want to analyze the case where a finite circular shell of null dust collapses in a de Sitter universe. To that end, we choose 
\bea\label{F}
F(v) =1 \hspace{2mm} &\mbox{if}& v<0; \nonumber \\
\dot F <0 \hspace{2mm} &\mbox{if}& 0\leq v < \bar v ;\\
F(v)=F(\bar v) =K ,\hspace{2mm} &\mbox{if}& v \geq \bar v \nonumber .
\eea
In general, these spacetimes have a portion of standard dS spacetime, given by the negative values of the advanced time $v$. In this portion $r=0$ is a regular centre of symmetry and the MTT is a standard null horizon placed at $r=\ell$. When this null horizon meets the shell at $v=0$, it becomes a timelike membrane. The future fate of this membrane then depends on the particular values of the final constant $K$. 

When the shell reaches the centre at $r=0$ it produces a curvature singularity. This curvature singularity always starts being timelike and it only becomes spacelike for negative values of $F(v)$, if they exist. Furthermore, this remains as a curvature singularity until the shell ends to flow in at $v =\bar v$. What happens then again depends on the value of $K$. 

Thus, let us analyze the different possibilities in turn.
\subsubsection{The case with $K=k^2 <1$}\label{conical}
\begin{figure}[h!]
\includegraphics[width=15cm]{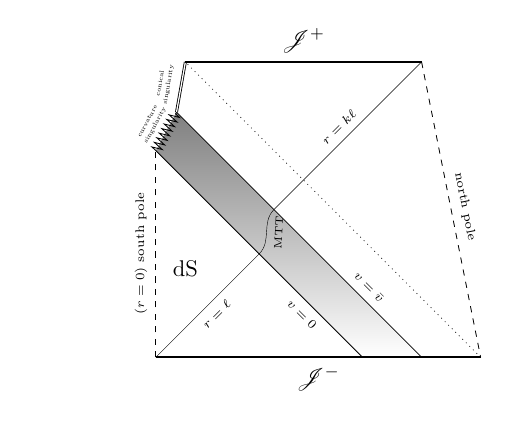}
\caption{\small{A conformal diagram of the collapse to a conical singularity. The shadowed region represents the null matter. There is an MTT which  is timelike within the matter, but otherwise null. These null portions of the MTT are represented by $r=\ell$ before the shell arrives and by $r=k\ell$ after it has passed, with $F(\bar v) =-k^2$. The curvature and conical singularities are drawn with a zigzag line, and a double line, respectively.} \label{fig:conical}}
\end{figure}

In this case the final value of $F(v)$ remains positive. Thus, on the one hand the curvature singularity is timelike everywhere, and on the other hand for $v\in (\bar v,\infty)$ we are in the situation considered in the previous section with a conical singularity. These two singularities `meet' at $v=\bar v$. This can be considered the standard situation where the collapsing shell produces a conical singularity due to having cut out part of the standard sphere $\mathbb{S}^2$ and identifying the borders of the missing piece. The existence of future infinity is not modified, though.
A conformal diagram showing all these features is shown in Fig. \ref{fig:conical}.

\subsubsection{The case with $K=-a^2 $}
\begin{figure}[h!]
\includegraphics[width=14cm]{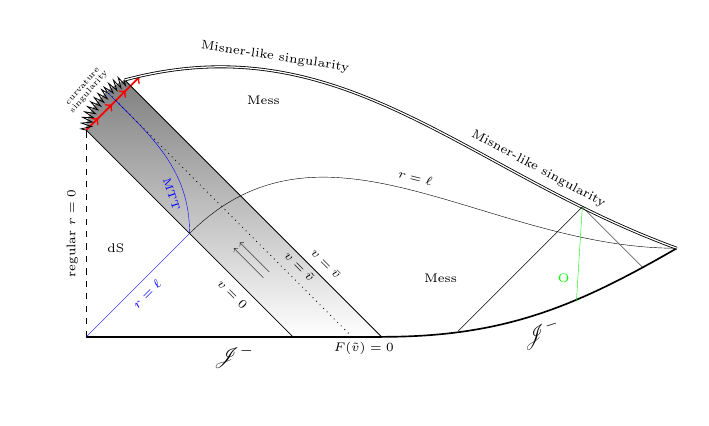}
\caption{\small{A conformal diagram for an ultramassive spacetime where the singularity is visible outside the collapsing shell. The matter shell is drawn as the shadowed zone.  The function $F(v)$ vanishes at $v=\tilde v$, and at that value the MTT disappears into the curvature singularity represented by a zigzag line. The outside region takes the form of a Mess spacetime \cite{Mess,A,BH} with a future Misner-type singularity marked by a double line. The red arrowed line indicates an outgoing null ray emitted from the curvature singularity that emerges into the Mess region, and thus it is visible for some outside observers before they arrive at the Misner-like singularity. The green line describes a possible observer who never sees the collapsing matter.}\label{fig:ultraM1}}
\end{figure}

\begin{figure}[h!]
\includegraphics[width=14cm]{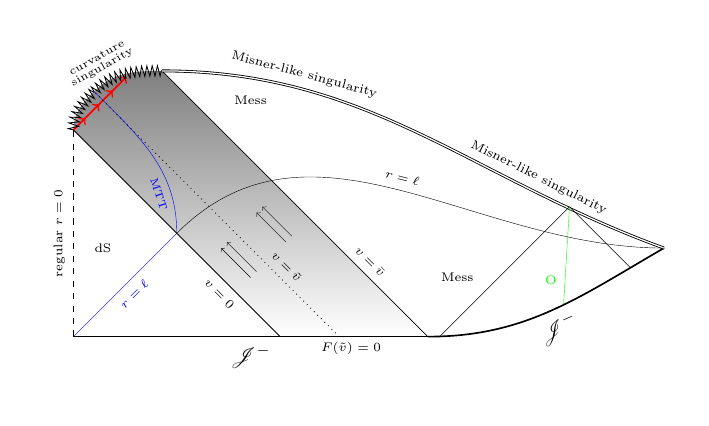}
\caption{\small{A conformal diagram for an ultramassive spacetime where the singularity is invisible outside the collapsing shell. Features and representations are the same as in Fig. \ref{fig:ultraM1} with the only difference that here the red arrowed line, representing a null ray emitted from the curvature singularity, never emerges outside the matter and ends up in another part of the curvature singularity.}\label{fig:ultraM2}}
\end{figure}

When the final value of the function $F$ is negative, the curvature singularity becomes spacelike for all values of $v$ such that $F<0$. The MTT within the shell is a timelike membrane that eventually disappears at the curvature singularity. This happens at the value of $v=\tilde v$ such that $F(v=\tilde v)=0$. At $v=\bar v$, when the null matter stops flowing in, the curvature singularity meets the Misner-like singularity which --together with the curvature singularity-- encloses the whole spacetime and becomes the final future fate for every possible particle in the spacetime. There is no future null infinity and there exist observers who will never see the matter nor the MTT, still they live in a contracting world and will end up at the future singularity. These characteristics are similar to those of the `ultra-massive' spacetimes in \cite{Snew}, and actually the Penrose diagrams look alike and have a handful of similar features as can be checked by comparison of Figs.\ \ref{fig:ultraM1} and \ref{fig:ultraM2} with Figure 2 in \cite{Snew}. Yet, there are two different possibilities in this situation, depending on whether or not the curvature singularity is visible {\em outside} the matter flow. 

To discuss these possibilities, we need to analyze the outgoing radial null geodesics inside matter. These are ruled by the ODE
\be\label{geod}
\frac{dr}{dv} = \frac{1}{2}(F(v)-\Lambda r^2) .
\ee
from where one immediately sees that there cannot be any such geodesics emanating from the singularity at $r=0$ for values of $v$ such that $F<0$ ---because the curvature singularity is spacelike there. At the beginning of the curvature singularity at $v=0$, i.e. when the shell first reaches the centre, we can choose the initial condition $r(0)=0$ for the outgoing radial null geodesic and one has 
$$
\left.\frac{dr}{dv} \right|_{v=0} = \frac{1}{2}.
$$
However, given that $F(v)$ is a non-increasing function of $v$ and it does decrease until it reaches negative values, there will always be a value $v=v_{{\rm{max}}}$ where 
$$
\left.\frac{dr}{dv} \right|_{v=v_{{\rm{max}}}} = 0
$$
defining the maximum value of $r$ reached by the geodesic: $r_{{\rm{max}}}:= r(v_{{\rm{max}}})$. Notice that $F(v=v_{{\rm{max}}})>0$ is necessarily positive, so this happens before $F$ becomes negative. From that value $v=v_{{\rm{max}}}$, $r$ starts to decrease and eventually will come back to vanish at, say, $v=v_{{\rm{end}}}$. The question is whether this will happen before or after the flow of incoming matter stops, that is to say, whether $v=v_{{\rm{end}}}$ is smaller or larger than $\bar v$. 

If $v_{{\rm{end}}} > \bar v$, then for all values of $v> \bar v$ the geodesic equation \eqref{geod} becomes simply
$$
\frac{dr}{dv} = -\frac{1}{2}(a^2 +\Lambda r^2) 
$$
whose solution is
$$
r(v)= a\ell \tan \left(\frac{a}{2\ell} (v_{{\rm{end}}}-v) \right) .
$$
In this case, the radial null ray emerging form the starting point of the curvature singularity eventually comes out of the matter shell before reaching the future Misner-like singularity. Therefore, there may be observers outside the shell that are able to see the curvature singularity before reaching the non-curvature one. This is depicted in Fig. \ref{fig:ultraM1}. 

If, on the other hand, $v_{{\rm{end}}} < \bar v$, then all rays coming out of the curvature singularity will end up in another portion of the curvature singularity and this singularity will never be visible outside the shell. This is shown in Fig. \ref{fig:ultraM2}.

In both cases, the spacetime outside the shell is isometric to the locally dS spacetimes with metric \eqref{metric1}. Therefore, these models are explicit examples of how to create Mess spacetimes \cite{Mess,A,BH} in 2+1 dimensions by collapsing matter.

\subsubsection{The critical case with $K=0$}\label{subsec:critical}
\begin{figure}[h!]
\includegraphics[width=14cm]{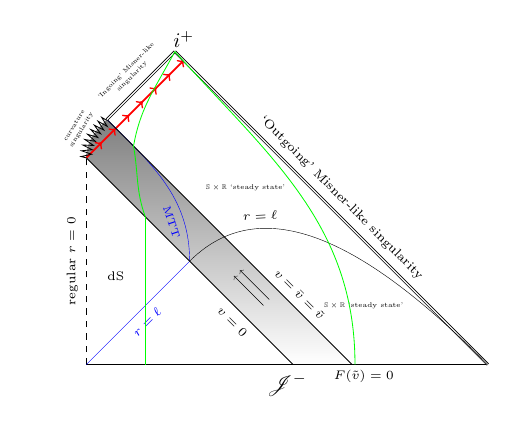}
\caption{\small{A conformal diagram of the critical case, where the Misner singularity is null and a portion of future infinity remains. Again the notation and representation are as in Figs.\ref{fig:ultraM1} and \ref{fig:ultraM2}, but now the Misner-like singularity is null and has two different portions, the one on the right where the outgoing null geodesics end, and that on the left where the ingoing null geodesics outside the matter end. In between, there is a remnant of future infinity, marked here by $i^+$, which some {\em complete} timelike geodesics  reach in infinite proper time. These are shown by the yellow lines. The region outside the shell is isometric to a `steady state' 2+1 spacetime with slices of topology $\mathbb{S}\times \mathbb{R}$.}\label{fig:Critical}}
\end{figure}

In this case the final value of the function $F$ vanishes: $F(\bar v)=0$. Again the timelike membrane inside the shell eventually disappears, in this case at the end of the curvature singularity and the beginning of the external, non-curvature, singularity. Now, the spacetime outside the shell is isometric to the steady-state version with cylindrical slices and metric \eqref{metric2}. As argued at the end of section \ref{dScases} this part of the spacetime contains a future singularity which is of Misner type for null geodesics with angular momentum, but it has two different parts, each representing the end of the outgoing and the ingoing null geodesics without angular momentum, respectively. These two portions are null and separated by a future timelike infinity towards which some complete timelike geodesics travel with no bound on proper time. The conformal diagram of the final spacetime is given in Fig. \ref{fig:Critical}. This can be compared with the extreme case of 4-dimensional ultra-massive spacetimes, Figure 4 in \cite{Snew}: the `point' P there, representing the remnants of future infinity, corresponds the the upper vertex separating the two null portions of the singularity in the critical case of Fig. \ref{fig:Critical}. 

\section{Discussion}\label{sec:discussion}
We have shown that in 2+1 spacetime dimensions with a positive cosmological constant, ultra-strong gravitating objects arise in a similar manner as in 3+1 dimensions. The so-called ultra-massive cases, objects more extreme than black holes \cite{Snew,Snew2}, also arise in a peculiar form in 2 + 1 dimensions, as shown in Figs. \ref{fig:ultraM1}, \ref{fig:ultraM2} and \ref{fig:Critical} and discussed in section \ref{sec:models}. Both the proper 4-dimensional ultra-massive case and the extreme case (having an extreme Kottler metric in the exterior, figure 4 in \cite{Snew}) are realized, with some particular different features, in 3-dimensional GR with $\Lambda >0$. The extreme case in \cite{Snew} would correspond to the critical case in this paper, see Fig. \ref{fig:Critical}.

The only ultra-strong gravity object that does not arise in 3-dimensional gravity with $\Lambda >0$ is the black hole. Of course this was well known, but the fact that the other, more extreme cases, do arise brings up the question as to why the black holes are missing. The situation similar to the formation of a black hole in 4 dimensions (figure 1 in \cite{Snew}) would be the case with $F(v)$ positive everywhere studied in subsection \ref{conical}. However, as we see in Fig. \ref{fig:conical}, the 2+1 model leads to a conical, naked, singularity rather than to a black hole. One may wonder what the reason might be for this difference. A possible answer is that, actually, in 2+1 gravity satisfying DEC there are no {\em dynamical horizons}, that is, spacelike MTTs,  as proven in \cite{Ida}, see also \cite{GSW}.

Such an absence of dynamical horizons is due to the instability {\em in spatial directions} of all possible marginally trapped circles. In what follows we provide a transparent argument
showing that there can certainly be MTTs ---as shown in Figs.\ref{fig:conical}--\ref{fig:Critical}---, but they will never form a dynamical horizon. Let us consider the stability operator for marginally trapped co-dimension 2 submanifolds, as introduced in \cite{AMS,AMS1}, see also section 2 in \cite{Snew2}. The main reason why the ultra-massive cases arise in 4 dimensions is the upper bound for the area of {\em spatially stable} marginally trapped spheres, and ultimately this bound appears due to the stability property.  This is clear from the application of the stability operator, as we explain next.

Any co-dimension two {\em marginally trapped} submanifold $S$ has a vanishing expansion $\theta_k$ along one normal null direction $k^\mu$, meaning that $\theta_k:= H_\mu k^\mu =0$. However, if we perturb the submanifold along an arbitrary vector field $n^\mu$ normal to $S$ (normalized $k^\mu n_\mu =1$), the corresponding perturbed mean curvature vector field loses (in general) its null character and acquires, at first order, a non-zero expansion given by
\be
\delta_{f\vec n} \theta^k:=L_n f
\label{deltatheta}
\ee
where $f$ is a function on $S$ providing the ``amount'' of the perturbation along $\vec n$ at each point on $S$. $L_n$ is called  the {\em stability operator} for $S$ in the direction $\vec n$ and is explicitly given by  \cite{AMS,AMS1}
\be
L_n f =-\Delta f+2s^B D_{B}f+f\left(\frac{\bar R}{2}-s^B s_{B}+D_{B}s^B-G_{\mu\nu}k^\mu \ell^{\nu}-\frac{n^\rho n_{\rho}}{2}\,  W\right) \label{Ln}
\ee
where $\bar R$ is the scalar curvature on $S$, $D_B$ its covariant derivative, $\Delta =D_B D^B$ its Laplacian,  $\ell^\mu$ the other independent null vector field normal to $S$ (normalized $\ell^\mu k_\mu =-1$), $s_{B}$ the one-form on $S$ defined by
$$
\bm{s} (\vec v) =k_{\mu}\nabla_{\vec v}\ell^\mu, \hspace{1cm} \mbox{for any vector field $\vec v$ tangent to $S$},
$$
and finally
$$
W:= G_{\mu\nu} k^\mu k^\nu + \Sigma_k^2
$$
with $\Sigma_k$ the shear scalar of $k^\mu$. Observe that, if the Einstein field equations and the null convergence condition hold, $W$ is non-negative.

$L_n$ is a (generally non-self-adjoint) elliptic operator on $S$. Nevertheless, as proven in \cite{AMS,AMS1}, it possesses a {\em real} principal eigenvalue denoted by $\lambda_n$. The corresponding {\em real} eigenfunction, denoted by $\phi_n$, can be chosen to be positive on all of $S$. A non-negative $\lambda_n$ means stability of the marginally trapped $S$ along the direction $n^\mu$.

For positive $f>0$, an equivalent way of writing \eqref{Ln} is \cite{Simon,Snew2}
\be
\frac{L_n f}{f} = -\Delta \ln f +D_B s^B-\left(D_B \ln f -s_B\right)\left(D^B \ln f -s^B\right)+\frac{\bar R}{2}-\frac{n^\rho n_{\rho}}{2}W-G_{\mu\nu}k^\mu \ell^{\nu} \label{Ln2}
\ee
which, applied to $f=\phi_n$ and using $L_n \phi_n = \lambda_n \phi_n$,  leads {\em in 3+1 dimensions}  to formula (16) in \cite{Snew2}
$$
\left(\Lambda +\lambda_n \right)A_S =4\pi -\int_S\left[\left(D_B \ln \phi_n -s_B\right)\left(D^B \ln \phi_n -s^B\right)+ \frac{n^\rho n_{\rho}}{2}W+T_{\mu\nu}k^\mu \ell^{\nu}\right]
$$
where one has used the Gauss-Bonnet theorem when integrating the scalar curvature on the marginally trapped sphere $S$. Here, $A_S$ denotes the area of the sphere. It must be noticed that, if DEC holds and $n^\mu$ is spacelike, all the terms in the integrand are non-negative and thus $(\Lambda +\lambda_n) \leq 4\pi$ leading to the area bound for marginally trapped spheres that are stable in the spatial direction $n^\mu$.

Compare this with the same formula in 2+1 dimensions. Given that the scalar curvature of any circle vanishes identically, in this case the corresponding formula for any marginally trapped circle ${\cal C}$ does not have any $4\pi$ term coming from the Gauss-Bonnet theorem and reads
$$
\left(\Lambda +\lambda_n \right)L_{\cal C}=-\int_{\cal C}\left[\left(D_B \ln \phi_n -s_B\right)\left(D^B \ln \phi_n -s^B\right)+ \frac{n^\rho n_{\rho}}{2}W+T_{\mu\nu}k^\mu \ell^{\nu}\right]
$$
where now $L_{\cal C}$ is the length of ${\cal C}$. In this case, if $\Lambda >0$ and DEC holds, then $\lambda_n<0$ is necessarily negative for any spacelike $n^\mu$. In other words, all marginally trapped circles are unstable in spatial directions. Therefore, marginally trapped circles perturbed outwardly {\em within a spacelike slice} inevitably lead to trapped circles. This implies that any MTT cannot be spacelike and, a fortiori, leads to the impossibility of having an event horizon for future null infinity. Thus, only conical (and naked) singularities may arise if future null infinity remains. If not, the ultra-massive cases (and the critical case) do emerge, and turn out to provide a natural `home' for the locally de Sitter spacetimes found by Mess. 

\section*{Acknowledgments}
J.M.M.S. is supported by the Basque Government grant number IT1628-22, and by Grant PID2021-123226NB-I00 funded by the Spanish MCIN/AEI/10.13039/501100011033 together with ``ERDF A way of making Europe'' .

\end{document}